\documentstyle[aps,pra,epsfig,twocolumn]{revtex} 
\def\bma{\begin{mathletters}}
\def\ema{\end{mathletters}}

\begin{document}

\title{Quasiparticle excitations and hierarchies of 4-dimensional quantum Hall fluid states in the matrix models}

\author{Yi-Xin Chen } 
\address{Zhejiang Institute of Modern Physics, Zhejiang University, 
             Hangzhou 310027, P. R.China}

\date{\today}
\maketitle

\begin{abstract}

We investigate the condensate mechanism of the low-lying excitations in the matrix models of 4-dimensional quantum Hall fluids recently proposed by us. It is shown that there exist some hierarchies of 4-dimensional quantum Hall fluid states in the matrix models, and they are similar to the Haldane's hierarchy in the 2-dimensional quantum Hall fluids. However, these hierarchical fluid states appear consistently in our matrix models without any requirement of modifications of the matrix models.


\vspace{.3cm}


\end{abstract}

\setcounter{equation}{0}

\indent

Recently, Susskind 
\cite{Susskind} proposed a description of quantum Hall effect (QHE) on the plane\cite{Laughlin} in terms of a non-commutative $U(1)$ Chern-Simons theory. Susskind's non-commutative 
Chern-Simons theory on the plane describes a spatially infinite quantum 
Hall system. The fields of this theory are infinite matrices that act on 
an infinite Hilbert space, appropriate to account for an infinite number of 
electrons. Polychronakos \cite{Polychronakos} proposed a 
matrix regularized version of Susskind's non-commutative Chern-Simons 
theory in effort to describe finite systems with a finite number of 
electrons in the limited spatial extent. This matrix model was shown to 
reproduce the basic properties of the quantum Hall fluid (QHF)\cite{Hellerman}. 

In the four dimensions, Zhang and Hu \cite{Zhang,Hu} have found a generalization of the quantum Hall effect on $S^4$. Elvang and Polchinski \cite{Elvang} have recently formulated the $SU(2)$ quantum Hall system on $R^4$, as the limit of Zhang' and Hu's that in the limit of $I\rightarrow \infty$. The authors in \cite{Fabinger,Chen,Karabali1,Kimura}have developed the idea of Zhang and Hu in other directions. In my recent paper\cite{Chen1}, I have established the effective description of non-commutative non-abelian Chern-Simons theories for the 4-dimensional QHF on the space of quaternions and Zhang-Hu's 4-dimensional QHF on $S^4$.

It is well known that the 2-dimensional QHF states with filling factor that are not of Laughlin type can be described by the hierarchical QHF states in Haldane's hierarchy\cite{Haldane} or in Jain's hierarchy\cite{Jain}. The hierarchical QHF states can be described by a (2+1)-dimensional gauge system of several $U(1)$ coupled Chern-Simons fields\cite{Wen}. The non-commutative generalization of such gauge system was considered in \cite{Rhalami,Rhalami1}.

Obviously, one want to ask whether the hierarchy of the QHF states exists in th four dimensions or not. If it does, how does one describe it? The goal of this paper is to answer the above problems. We shall show that there exist consistently the corresponding hierarchies of the 4-dimensional QHF states in our matrix models, and they can be built up by means of the Fock basis' of the matrix models.

Let us first recall the matrix models\cite{Chen1} of the 4-dimensional QHFs proposed by us. The 4-dimensional QHF on the space of quaternions is described by the matrix model with the action
\begin{eqnarray}
S&=&S[Z^{\alpha}, \Psi^{\alpha}, A_{0}] \nonumber\\
&=&\frac{H}{4}\int dt Tr\{ {\bar Z}^{\alpha} i(\partial_t Z^{\alpha} +[A_{0}^{\alpha\beta},Z^{\beta}])- 2i\theta\sigma_{3}A_{0} \nonumber\\
&-&\omega {\bar Z}^{\alpha}Z^{\alpha} \}
+\frac{1}{2}\int dt\Psi^{\dagger\alpha}i(\partial_t \Psi^{\alpha} +A_{0}^{\alpha\beta}\Psi^{\beta})+ h.c.
\end{eqnarray}
Here, we consider the system composed of $N$ particles. The complex coordinates $Z^{\alpha}$, which can be expressed as 
$Z^{1}=Q^1+iQ^2$ and $Z^{2}=Q^3+iQ^4$ in terms of four real coordinates, are represented by finite $N\times N$ matrices. The 
$2\times 2$ matrix built up by the gauge fields $A_{0}^{\alpha\beta}$ with the indices $(\alpha,\beta)$  is anti-hermitian and traceless except for that each element of them is a $N\times N$ matrix. The constant $H$ plays the role similar to the constant magnetic field in 2-dimensional QHE. $\theta$ is the positive parameter characterizing the non-commutativity of the coordinates of particles, and $\sigma_{3}$ the third component of Pauli matrix. After the quantization 
of the system, the matrix elements $Z^{\alpha}_{ij}$,  $Z^{\alpha\dagger}_{ij}$ and the components 
$\Psi^{\alpha}_{i}$,  $\Psi^{\alpha\dagger}_{i}$ become the operators satisfy the canonical commutation relations. Based on the construction of solutions obeying the constraint equations of the physical states, we have determined the Fock basis of the physical states of the matrix model as
\begin{eqnarray}
|\{ c_{i} \} ,&\{ &c_{i}^{\prime} \}, k \rangle =\prod_{j=1}^{N}(Tr Z^{1\dagger j})^{c_{j}}(Tr Z^{2\dagger j})^{c_{j}^{\prime}} \nonumber\\
&(&\epsilon^{i_{1}\cdots i_{N}}\prod_{n=1}^{N}(\Psi^{1\dagger}Z^{1\dagger N-n}Z^{2\dagger n-1})_{i_{n}}
)^{k}|0 \rangle  ,
\end{eqnarray}
where, $\sum_{i=1}^{N}i(c_{i} -c_{i}^{\prime} )=lN$, and $l=0,1, \cdots, k$.

In order to establish the effective description of non-commutative field theory for Zhang' and Hu's 4-dimensional QHF  on $S^4$, we introduced the action of $SO(4)$ matrix Chern-Simons model\cite{Chen1} as following
\begin{equation}
S_{M} =S[Z^{\alpha}, \Psi^{\alpha}, A_{0}]+S[Z^{{\bar \alpha}}, \Psi^{{\bar \alpha}}, {\bar A}_{0}]  
\end{equation}
with the geometrically restricted condition $Tr({\bar Z}^{a}Z^{a}) =G$, where $G$ is a parameter dependent on the model. The Fock basis of physical states of this $SO(4)$ matrix Chern-Simons model are given by
\begin{eqnarray}
| &\{&c^a_i\},k;I \rangle 
=\prod_{j=1}^{N}(Tr Z^{1\dagger j})^{c^1_{j}}(Tr Z^{2\dagger j})^{c^2_{j}}\nonumber\\
&(&Tr Z^{3\dagger j})^{c^3_{j}}(Tr Z^{4\dagger j})^{c^4_{j}}
(\epsilon^{i_{1}\cdots i_{N=d_{(2I,0)_{5}}}} \prod_{n=0}^{I}\nonumber\\
&[&\Psi^{a\dagger}Z^{a\dagger}]^{(I-n,n)_{4}}_{i_{\sum_{m=1}^{n-1}d_{(I-m,m)_{4}}+1}\cdots i_{\sum_{m=1}^{n}d_{(I-m,m)_{4}}}})^{k}|0 \rangle ,
\end{eqnarray}
where, $\sum_{i=1}^{N}i(c_{i}^{1}-c_{i}^{2})=l_{1}(2j_{1}+1)$ for $l_{1}=0,1,\cdots, k$ and $\sum_{i=1}^{N}i(c_{i}^{3}-c_{i}^{4})=l_{2}(2j_{2}+1)$ for $l_{2}=0,1,\cdots, k$, in which $j_{1}+j_{2}=I, 2j_{1}=0,1,\cdots, 2I$. The number $N$ of particles is equal to the dimensionality $d_{(2I,0)_5}$ of the $SO(5)$ irrep. $(2I,0)_5$. The 
elements $[\Psi^{a\dagger}Z^{a\dagger}]^{(j_{1},j_{2})_{4}}_{i_{1}\cdots i_{d_{(j_{1},j_{2})_{4}}}}$ in (4) are defined by
\begin{eqnarray}
[\Psi^{a\dagger}Z^{a\dagger}]^{(j_{1},j_{2})_{4}}_{i_{1}\cdots i_{d_{(j_{1},j_{2})_{4}}}}
&=& \prod_{n=0,m=0}^{2j_{1},2j_{2}}(\Psi^{1\dagger}Z^{1\dagger 2j_{1}-n}Z^{2\dagger n}\nonumber\\
&\Psi^{3\dagger}&Z^{3\dagger 2j_{2}-m}Z^{4\dagger m})_{i_{(n+1)(m+1)}} .
\end{eqnarray}

As pointed by Polychronakos\cite{Polychronakos1}, the classical value of the inverse filling fraction is shifted quantum mechanically if one use the finite matrix Chern-Simons theory to describe the fractional quantum Hall states. This can be equivalently viewed as a renormalization of the Chern-Simons coefficient. In fact, this level shift of the matrix Chern-Simons model can be read off from the well known quantum mechanically level shift of the corresponding Chern-Simon theory. The renormalization of the level of Chern-Simons theory has been finished by using a biparameter family of BRS invariant regularization methods of Chern-Simons theory \cite{Giavarini}. This renormalization leads to the level shift being $k\rightarrow k+sign(k)c_{V}$, where $k$ is the bare Chern-Simons level parameter and $c_{V}$ the quadratic Casimir operator in the adjoint representation of the gauge group of Chern-Simons theory. This implies that the $SU(2)$ Chern-Simons matrix model at level $k$ should be identified with the 4-dimensional quantum Hall states on the space of quaternions at the filling fraction $1/(k+2)$, rather than $1/k$. That is, the Laughlin type wavefunction of the 4-dimensional QHF for filling fraction $1/(k+2)$ can be equivalently described by the physical ground state of the $SU(2)$ Chern-Simons matrix model
\begin{equation} 
|\{ 0\} ,\{ 0 \}, k \rangle =
(\epsilon^{i_{1}\cdots i_{N}}\prod_{n=1}^{N} 
(\Psi^{1\dagger}Z^{1\dagger N-n}Z^{2\dagger n-1})_{i_{n}}
)^{k}|0 \rangle .
\end{equation}

The description of Zhang' and Hu's 4-dimensional QHF on $S^4$ is prvovided by the $SO(4)$ Chern-Simons matrix model (3) in which $S^7$ is from the second fibration of $S^4$. The action (3) is the sum of two $SU(2)$ Chern-Simons matrix actions. The physical quantum states of this model are given by all admissible $SO(4)$ blocks, which sre built up by the direct products of two $SU(2)$ fundamental blocks. So the filling fraction of Zhang' and Hu's 4-dimensional quantum Hall states is also shifted to $k+2$ if one use the $SO(4)$ Chern-Simons matrix model at the level $k$ to describe the 4-dimensional QHF on $S^4$. This means that the quantum Hall states on $S^4$ at filling fraction $1/(k+2)$ should be identified with the $SO(4)$ Chern-Simons matrix model at the level $k$. The Laughlin type wavefunction of this QHF at filling fraction $1/(k+2)$ corresponds to the physical ground state of the $SO(4)$ matrix model
\begin{eqnarray}
| &\{&c^a_i =0\},k;I \rangle =
(\epsilon^{i_{1}\cdots i_{N=d_{(2I,0)_{5}}}} \prod_{n=0}^{I}\nonumber\\
&[&\Psi^{a\dagger}Z^{a\dagger}]^{(I-n,n)_{4}}_{i_{\sum_{m=1}^{n-1}d_{(I-m,m)_{4}}+1}\cdots i_{\sum_{m=1}^{n}d_{(I-m,m)_{4}}}})^{k}|0 \rangle ,
\end{eqnarray}
where, $\sum_{i=1}^{N}i(c_{i}^{1}-c_{i}^{2})=l_{1}(2j_{1}+1)$ for $l_{1}=0,1,\cdots, k$ and $\sum_{i=1}^{N}i(c_{i}^{3}-c_{i}^{4})=l_{2}(2j_{2}+1)$ for $l_{2}=0,1,\cdots, k$, in which $j_{1}+j_{2}=I, 2j_{1}=0,1,\cdots, 2I$.

The low-lying excitations in our matrix models can be described in terms of quasiparticles and quasiholes \cite{Polychronakos}. A quasiparticle state is obtained by peeling a 'particle' from the surface of the Fermi sea. That is, one quasiparticle obtained by exciting a 'particle' at Fermi level by energy amount $n\omega$ is described by
\begin{eqnarray}
p^{1\dagger}_{n}|\{ 0&\}& ,\{ 0 \}, k \rangle =
(\epsilon^{i_{1}\cdots i_{N}}\prod_{m=1}^{N} 
(\Psi^{1\dagger}Z^{1\dagger N-m} \nonumber\\
Z^{2\dagger m-1}&)&_{i_{m}}
)^{k-1} 
\epsilon^{i_{1}\cdots i_{N}} 
(\Psi^{1\dagger}Z^{1\dagger N-1+n}Z^{2\dagger 0})_{i_{1}} \nonumber\\
\prod_{m=2}^{N}&(&\Psi^{1\dagger}Z^{1\dagger N-m}Z^{2\dagger m-1})_{i_{m}}|0\rangle .
\end{eqnarray}
The quasiholes correspond to the minimal excitations of the ground state inside the QHF. One quasihole excitation is obtained by creating a gap inside the QHF with the energy increase $m\omega$
\begin{eqnarray}
h^{1\dagger}_{m}|\{ 0&\}& ,\{ 0 \}, k \rangle =
(\epsilon^{i_{1}\cdots i_{N}}\prod_{n=1}^{N} 
(\Psi^{1\dagger}Z^{1\dagger N-n} \nonumber\\
Z^{2\dagger n-1}&)&_{i_{n}}
)^{k-1} 
\epsilon^{i_{1}\cdots i_{N}}\prod_{n=1}^{m}
(\Psi^{1\dagger}Z^{1\dagger N-n+1}Z^{2\dagger n-1})_{i_{n}} \nonumber\\
\prod_{n=m+1}^{N}&(&\Psi^{1\dagger}Z^{1\dagger N-n}Z^{2\dagger n-1})_{i_{n}}|0\rangle .
\end{eqnarray}
Obviously, $p^{1\dagger}_{1}=h^{1\dagger}_{1}$. So there is no fundamental distinction between 'particles' and 'holes' in the matrix model. Similarly, one can describe the quasiparticle $p^{2\dagger}_{n}$ and quasihole $h^{2\dagger}_{m}$ of excitations corresponding to the oscillator field $Z^{2}$. Although all of these excitations are the fundamental excitations in the matrix Chern-Simons model, however, they can not be regarded directly as the physical low-lying excitations in the matrix model. The physical exciting states must obey the constraint condition of physical states of the matrix model. By using of the techniques in \cite{Hellerman,Chen1}, one can find that all of the fundamental excitations as mentioned above can be expressed by the following states
\begin{eqnarray}
&P_{n_1,n_2}^{\dagger}&|\{ 0\} ,\{ 0 \}, k \rangle = \nonumber\\
&\prod_{j=1}^{N}&(Tr Z^{1\dagger j})^{c_{j}}(Tr Z^{2\dagger j})^{c_{j}^{\prime}} 
|\{ 0\} ,\{ 0 \}, k \rangle ,
\end{eqnarray}
where, $n_1=\sum_{i=1}^{N}ic_{i}$ and $n_2=\sum_{i=1}^{N}ic_{i}^{\prime}$. The quasiparticle excitations associated with the oscillator field $Z^1$ are more than those associated with the field $Z^2$ if $n_1$ is larger than $n_2$. If $n_1 =n_2 +lN$ for $l=1,\cdots,k$, $P_{n_1,n_2}^{\dagger}|\{ 0\} ,\{ 0 \}, k \rangle $ describe the physical excitations of the matrix model. The existence of hierarchy of the 4-dimensional QHF states is just due to the condesation of these physical excitations.

If we denote the quantum number of the $SU(2)$ representation as $J$ and noticing $Z^{1\dagger}$ and $Z^{2\dagger}$ associated with 'spin up' and 'spin down' respectively,  we can find that the number of particles $N=2J+1$ for the case of full filling. That is, the particles of this 4-dimensional QHF are filled in the representation of the $SU(2)$ level $J$. 
For the case of general ${\tilde k}=k+2$ from the level shifting, the degeneracy given by the representation of $SU(2)$ is $2J^{\prime} +1={\tilde k}(N-1)+1$, while the particle number is still $N$. The filling factor in this case is $\nu =N/N{\tilde k}=1/{\tilde k}$ in the thermodynamic limit. Hence, the state $|\{ 0\} ,\{ 0 \}, k \rangle $ is equivalent to the Laughlin type wavefunction of the 4-dimensional QHF on the space of quaternions for the filling fraction $1/{\tilde k}=1/(k+2)$. The particles are filled in the degenerate states fixed by the representation of the $SU(2)$ level ${\tilde k}$. So the condition of the physical states  $n_1 =n_2 +lN$ for $l=1,\cdots,k$ implies that these states describe the collective excitations with the total 'spin' $lN/2$ of the primary 4-dimensional QHF described by the state $|\{ 0\} ,\{ 0 \}, k \rangle $. Following Haldane\cite{Haldane}, we can construct the collective ground state of the 4-dimensional excitation fluid by the condensing of the quasiparticle and quasihole excitations. The Laughlin-type fluid state including the excitations is given by
\begin{eqnarray}
&|\{0\}&,\{0\}, p_1, k \rangle =
(\epsilon^{i_1,\cdots,i_{N_1}} \nonumber\\
&\prod_{n=1}^{N_1}&(P^{1\dagger N_1 -n}P^{2\dagger n-1})_{i_n}))^{p_1}
|\{ 0\} ,\{ 0 \}, k \rangle ,
\end{eqnarray}
where $N_1=N/p_1 +1$ and $N$ is divisible by $p_1$. $P^{1\dagger i}$ and $P^{2\dagger i}$ represent $P_{i,0}^{\dagger}$ and $P_{0,i}^{\dagger}$ respectively. One can construct the excitation states of the 4-dimensional excitation fluid in the similar way of the construction of the excitation states in the 4-dimensional QHF. The result is
\begin{eqnarray}
|&\{c_{1i}\}&,\{c_{1i}^{\prime}\}, p_1, k \rangle = \nonumber\\
&\prod_{j=1}^{N_1}&(P_{1}^{1\dagger j})^{c_{1j}}(P_{1}^{2\dagger j})^{c_{1j}^{\prime}}|\{0\},\{0\}, p_1, k \rangle ,
\end{eqnarray}
where, $P_{1}^{\alpha\dagger j}=\sum_{i=1}^{N_1}P_{i}^{\alpha\dagger j}$. Although these states do not satisfy generally all of constraint conditions of the physical states in the matrix model, they are the mediate states to construct the ground state of the 4-dimensional excitation fluid obeying the constraint conditions. 

Similar to the 2-dimensional QHFs, the procedure of constructing the 4-dimensional excitation fluids can be iterated, and leads to the hierarchy of the 4-dimensional QHF states. We give the result of the iterated construction of the QHF states as following
\begin{eqnarray}
&|&\{0\},\{0\}, p_{m},\cdots,p_1, k \rangle =
\prod_{q=1}^{m}(\epsilon^{i_1,\cdots,i_{N_q}} \nonumber\\
&~&\prod_{n=1}^{N_q}(P_{q-1}^{1\dagger N_q -n}P_{q-1}^{2\dagger n-1})_{i_{n}})^{p_q}|\{ 0\} ,\{ 0 \}, k \rangle ,
\end{eqnarray}
where $p_q(N_q-1)+N_{q+1}=N_{q-1}$ with $N_{q}=0$ for $q>m$ and $N_{0}=N$. Remember that the degeneracy of the 4-dimensional QHF states is determined by the level of the $SU(2)$ representation. For the hierarchical states of the 4-dimensional QHF, the level is given by $\frac{1}{2}{\tilde k}(N-1)+\frac{1}{2}N_1$. Solving the iteration relations of the hierarchical states, we obtain the filling factor of the m-th hierarchical fluid state in the thermodynamic limit
\begin{equation}
\nu=\frac{1}{{\tilde k}+\frac{1}{p_1+\frac{1}{p_2+\cdots+\frac{1}{p_m}}}}.
\end{equation}
This has clearly shown Haldane's hierarchical structure in the matrix model of the 4-dimensional QHF. However, (13) describes the hierarchical fluid states from the condensation of the excitations for the case of $l=1$ in the physical states (2). In fact, there exist the corresponding hierarchical fluid states for each $l$. Based on the above discussion, one can easily write off these hierarchical states. The result is
\begin{eqnarray} 
|&\{&0\},\{0\}, p_{m}^l,\cdots,p_1^l, l, k \rangle =
\prod_{q=1}^{m}(\epsilon^{i_1,\cdots,i_{N_q}} \nonumber\\
&~&\prod_{n=1}^{N_q}(P_{q-1}^{1\dagger N_q -n}P_{q-1}^{2\dagger n-1})_{i_{n}})^{p_q^l}|\{ 0\} ,\{ 0 \}, k \rangle ,
\end{eqnarray}
where $p_q^l(N_q-1)+lN_{q+1}=lN_{q-1}$ with $N_{q}=0$ for $q>m$ and $N_{0}=N$. The filling factors corresponding to the states (15) are
\begin{equation}
\nu=\frac{1}{{\tilde k}+\frac{l}{p_1^l+\frac{l}{p_2^l+\cdots+\frac{l}{p_m^l}}}}
\end{equation}
for $l=1,2,\cdots,k$. It should be emphasized that we have exihibited the rich hierarchical structures of the 4-dimensional QHF by means of the fundamental fields $Z^{\alpha}$ and $\Psi^{\alpha}$ of the matrix model rather than adding to extra fields. This implies that the matrix model not only describes the 4-dimensional Laughlin type QHF but also does the hierarchy of the excitation fluids derived from it.

As mentioned above, the physical quantum states of the $SO(4)$ Chern-Simons matrix model are given in terms of all admissible $SO(4)$ blocks built up by the direct products of two $SU(2)$ fundamental blocks. So the collective excitations of the 4-dimensional QHF described by the matrix model must be built up in the same way. The exciting states of quasiparticles and quasiholes in this model are given by
\begin{equation}
P_{n_1,n_2,n_3,n_4}^{\dagger}|\{ 0\},I, k \rangle =
\prod_{a=1}^{4}\prod_{j=1}^{N}(Tr Z^{a\dagger j})^{c_{j}^a} 
|\{ 0\},I, k \rangle ,
\end{equation}
where, $n_a=\sum_{i=1}^{N}ic_{i}^a$. Now we can use the result of the $SU(2)$ Chern-Simons matrix model to construct the hierarchical fluid states from the condensation of the excitations in the present matrix model. In order to do this, we denote $P_{i,0,0,0}^{\dagger}$ and $P_{0,i,0,0}^{\dagger}$ as $P^{1\dagger i}$ and $P^{2\dagger i}$ respectively. Thus $P^{1\dagger i}$ and $P^{2\dagger i}$ describe the fundamental excitations of the 4-dimensional Laughlin type QHF associated with one $SU(2)$ of $SO(4)$. Those associated with the other $SU(2)$ of $SO(4)$ are described by $P^{3\dagger i}=P_{0,0,i,0}^{\dagger}$ and $P^{4\dagger i}=P_{0,0,0,i}^{\dagger}$. The excitation condensation of Zhang' and Hu's 4-dimensional QHF on $S^4$ results in the 4-dimensional excitation fluid. The ground state of it is read as
\begin{eqnarray}
&|&\{0\},p_1,I, k \rangle =
(\prod_{N_1^1,N_1^2}\epsilon^{i_1,\cdots,i_{N_1^1\times N^2_1}} 
\prod_{n=1,m=1}^{N_1^1,N_1^2}
(P^{1\dagger N_1^1 -n}\nonumber\\
&~&P^{2\dagger n-1}P^{3\dagger N_1^2 -m}P^{4\dagger m-1})_{i_{n\times m}})^{p_1}
|\{0\},I, k \rangle ,
\end{eqnarray}
where $N_1^1=N^1/p_1+1=(2j_1+1)/p_1+1$ and $N_1^2=N^2/p_1+1=(2j_2+1)/p_1+1$. Since the particles or quasiparticles of the matrix model are filled in the representation of $SU(2)\times SU(2)$, i.e., $SO(4)$, we can calculate the number of the extra quasiparticles according to the total 'spin' contributed by these quasiparticles\cite{Haldane}. So $N^{ex}/4=\sum_{N_1^1,N_1^2}\frac{1}{2}(N_1^1-1)\times \frac{1}{2}(N_1^2-1)=\frac{1}{4}\sum_{j_1,j_2}\frac{2j_1+1}{p_1}\times \frac{2j_2+1}{p_1}$ for $j_1+j_2=I$ and $2j_1=0,1,\cdots,2I$. Noticing both $N^1=2j_1+1$ and $N^2=2j_2+1$ should be divisible by $p_1$, we finish the above sum to get $N^{ex}=d_{(2I/p_1,0)_5}$. 

Paralleling with the construction of the hierarchical fluid states of the $SU(2)$ matrix model, the hierarchical fluid states of Zhang' and Hu's 4-dimensional QHF can be constructed as the following 
\begin{eqnarray}
&|&\{0\},p_{m},\cdots,p_1, k \rangle =
\prod_{q=1}^{m}\prod_{N_q^1,N_q^2}(\epsilon^{i_1,\cdots,i_{N_q^1\times N_q^2}}
\prod_{n=1,m=1}^{N_q^1,N_q^2} \nonumber\\
&(&P_{q-1}^{1\dagger N_q^1 -n}P_{q-1}^{2\dagger n-1}P_{q-1}^{3\dagger N_q^2 -m}P_{q-1}^{4\dagger m-1})_{i_{n\times m}})^{p_q}|\{ 0\},I,k \rangle ,
\end{eqnarray}
where $p_q(N_q^i-1)+N_{q+1}^i=N_{q-1}^i$ with $N_{q}^i=0$ for $q>m$ and $N_{0}^i=2j_i+1$, $i=1,2$. The excitation operators $P_{q}^{a\dagger}$ in (19) are given by
\begin{equation}
|\{c_{qi}^a\},p_q,\cdots,p_1,I,k \rangle =
\prod_{a=1}^{4}\prod_{j=1}^{N_q}(P_{q}^{a\dagger j})^{c_{qj}^a}|\{0\},p_q,\cdots,p_1,I,k \rangle ,
\end{equation}
where, $P_{q}^{a\dagger j}=\sum_{i=1}^{N_1}P_{qi}^{a\dagger j}$. From the solution of iteration relations, we can read off the filling factor of the 4-dimensional hierarchical fluid state on $S^4$. In the thermodynamic limit, which can be taken by setting $I\rightarrow \infty$ since the number of particles is equal to $N=d_{(2I,0)_5}=\frac{1}{6}(2I+1)(2I+2)(2I+3)$, the filling factor of the m-th hierarchical fluid state is
\begin{equation}
\nu=\frac{1}{{\tilde k}^3+(\frac{1}{p_1+\frac{1}{p_2+\cdots+\frac{1}{p_m}}})^3}.
\end{equation}

For the general $l$ in the solutions of the physical states, we can also construct the hierarchical fluid states from the condensation of the $l$-th excitations of the 4-dimensional Laughlin type QHF. They are given by
\begin{eqnarray} 
&|&\{0\},p_{m}^l,\cdots,p_1^l, l,I, k \rangle =
\prod_{q=1}^{m}\prod_{N_q^1,N_q^2}(\epsilon^{i_1,\cdots,i_{N_q^1\times N_q^2}} 
\prod_{n=1,m=1}^{N_q^1,N_q^2} \nonumber\\
&(&P_{q-1}^{1\dagger N_q^1 -n}P_{q-1}^{2\dagger n-1}P_{q-1}^{3\dagger N_q^2 -m}P_{q-1}^{4\dagger m-1})_{i_{n\times m}})^{p_q^l}|\{ 0\},I,k \rangle ,
\end{eqnarray}
where $p_q^l(N_q^i-1)+lN_{q+1}^i=lN_{q-1}^i$ with $N_{q}^i=0$ for $q>m$ and $N_{0}^i=2j_i+1$, $i=1,2$. Their filling factors are read as 
\begin{equation}
\nu=\frac{1}{{\tilde k}^3+(\frac{l}{p_1+\frac{l}{p_2+\cdots+\frac{l}{p_m}}})^3},
\end{equation}
for $l=1,2,\cdots,k$.

In summary, we have found the hierarchical structures of the 4-dimensional QHF states in the matrix models of 4-dimensional QHFs. The arguements of the renormalization of the Chern-Simons coefficient result in the conclusion that the classical values of the inverse filling fractions are shifted quantum mechanically by two units in the matrix models, i.e., $1/k\rightarrow 1/(k+2)$. We have given the constructions of the hierarchical states of the 4-dimensional QHFs (15) and (22) corresponding to the QHF of the space of quaternions and the QHF on $S^4$, respectively. This implies that our matrix models can effectively describe not only the 4-dimensional QHFs but also their hierarchies.\\

I would like to thank Dimitra Karabali for many valuable suggestions and comments. 
The work was partly supported by the NNSF of China (Grant No.90203003) and by the Foundation 
of Education Ministry of China (Grant No.010335025).


\end{document}